\documentclass[aps,amsmath,showpacs,amsfonts,10pt,twocolumn]{revtex4}
\usepackage{epsfig,graphicx}
\usepackage[english]{babel}
\usepackage{amsfonts}
\usepackage{amsmath}
\usepackage{latexsym}
\usepackage{graphics,bm}
\usepackage{natbib}
\usepackage{dcolumn}
\usepackage{bm}
\usepackage{rotating}

\begin{document}

\title{Non-Equilibrium Quantum Phases of Two-Atom Dicke Model}

\author{Aranya B. Bhattacherjee}

\address{School of Physical Sciences, Jawaharlal Nehru University, New Delhi-110067, India}

\begin{abstract}
In this paper, we investigate the non-equilibrium quantum phases of the two-atom Dicke model, which can be realized in a two species Bose-Einstein condensate interacting with a single light mode in an optical cavity. Apart from the usual non-equilibrium normal and inverted phases, a non-equilibrium mixed phase is possible which is a combination of normal and inverted phase. A new kind of quantum phase transition is predicted from non-superradiant mixed phase to the superradiant phase which can be achieved by tuning the two different atom-photon couplings. We also show that a quantum phase transition from the non-superradiant mixed phase to the superradiant phase is forbidden for certain values of the two atom-photon coupling strengths.

\end{abstract}


\maketitle

\section{Introduction}

The interaction of a collection of atoms with a radiation field has always been an important topic in quantum optics. The Dicke model (DM) which describes interaction of $N$ identical two level atoms with a single radiation field mode, established the importance of collective effects of atom-field interaction, where the intensity of the spontaneously emitted light is proportional to $N^{2}$ rather than $N$ \citep{dicke}. The spatial dimensions of the ensemble of atoms are smaller than the wavelength of the radiation field. As a result, all the atoms experience the same field and this gives rise to the collective and cooperative interaction between light and matter. The DM exhibits a second-order quantum phase transition (QPT) from a non-superradiant normal phase to a superradiant phase when the atom-field coupling constant exceeds a certain critical value \citep{hepp,wang,hioes,tobias}. The experimental observation of the QPT predicted in the DM required that the collective atom-photon coupling strength to be of the same order of magnitude as the energy separation between the two atomic levels. In conventional atom-cavity setup this condition was impossible to satisfy until it was observed recently in a trapped Bose-Einstein condensate (BEC) in an optical cavity \citep{bau1,bau2,brenn,ritsch}. In the BEC setup, the two spin states of the original DM are the two momentum states of the BEC which are controlled by the atomic recoil energy and Raman pumping schemes. This approach is similar to a novel scheme proposed by Dimer et. al.\citep{dimer}. An important aspect of these experimental developments is the possibility to explore exotic phases mediated by the cavity field. The superradiance phase transition in a BEC is accompanied by self-organization of the atoms into a checker board pattern \citep{bau1,bau2,brenn,ritsch,nagy}.

Interesting equilibrium and non-equilibrium phases have been predicted in the DM with BEC \citep{bhaseen,liu}, including crystallization and frustration \citep{gopal1}, as well as spin glass phase \citep{gopal2,strack,buch,And}. Multimode DM has also been explored recently, revealing interesting physics such as Abelian and non-Abelian gauge potentials \citep{larson1}, spin-orbit induced anomalous Hall effect \citep{larson2}, and prediction of the Nambu-Goldstone mode \citep{fan}. An interesting extension of the BEC Dicke model is the optomechanical Dicke model which has been proposed for detection of weak forces \citep{bhat1,bhat2}. In the present paper, we investigate the non-equilibrium properties of the two-atom Dicke model, which can be realized by two species BEC in an optical cavity. Apart from the usual non-equilibrium normal and inverted phases, the dynamical phase diagram reveals a new kind of non-equilibrium mixed phase. This gives rise to a new quantum phase transition from the mixed phase to the superradiant phase by manipulation of the two distinct atom-photon coupling strengths. In addition, we show that a quantum phase transition from the non-superradiant phase to the superradiant phase is not allowed for certain values of the atom-photon coupling strengths of the two set of atoms.

\section{The Model}

\begin{figure}[h]
\hspace{-0.0cm}
\includegraphics [scale=0.6]{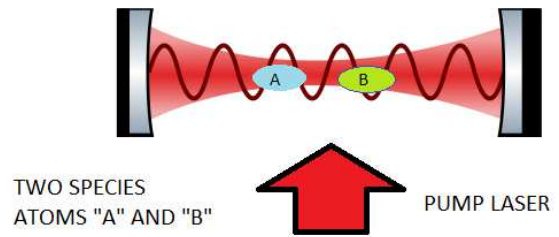}
\caption{(color online)Experimental setup showing two sets of cold atoms (blue and green) in an optical cavity with transverse pumping. The two sets of atoms have different atom-photon coupling strengths which depends on their position in the cavity. On increasing the transverse pump intensity, one type of atoms can reach the critical point earlier. }
\end{figure}\label{fig1}

We consider two different ensembles of $N_{1}$ and $N_{2}$ atoms coupled simultaneously to the quantized field of an optical cavity mode (Fig.1). The two sets of atoms have transition frequencies $\omega_{1}$ and $\omega_{2}$ while the frequency of the cavity mode is $\omega_{c}$. The cavity is pumped by an transverse external laser with frequency $\omega_{p}$. The light-matter coupling strengths for the two sets of atoms are $\lambda_{1}$ and $\lambda_{2}$. These coupling strengths $\lambda_{1}$ and $\lambda_{2}$ can be written as $\lambda_{i}=\lambda_{0i} \Omega_{P}/2(\omega_{p}-\omega_{i})$ ($i=1,2$), $\lambda_{0i}$ is the single atom-cavity mode coupling while $\Omega_{P}$ is the transverse pump beam Rabi frequency. The detuning $(\omega_{p}-\omega_{i})$ is considered to be large so as to avoid spontaneous emission. The effective Hamiltonian of the system takes the form of a two-atom Dicke model with

\begin{eqnarray}
H&=&\hbar \omega_{1} J_{1z}+\hbar \omega_{2} J_{2z} +\hbar \omega_{c} a^{\dagger} a \\ \nonumber
&+& \frac{\hbar \lambda_{1}}{\sqrt{N_{1}}} \left( J_{1+}+J_{1-} \right)\left( a+a^{\dagger}\right)\\ \nonumber
&+& \frac{\hbar \lambda_{2}}{\sqrt{N_{2}}} \left( J_{2+}+J_{2-} \right)\left( a+a^{\dagger}\right),
\end{eqnarray}

where $\vec{J_{i}}=(J_{ix},J_{iy},J_{iz})$ is the effective collective spin of length $N_{i}/2$ for the two sets of atom and $J_{i \pm}=J_{ix}\pm i J_{iy}$.

We now discuss the non-equilibrium dynamics arising from the above two-atom Dicke model.The semi-classical equations of motion for the system are given by

\begin{equation}
\dot{J_{1z}}=\frac{i \lambda_{1}}{\sqrt{N_{1}}}\left(a^{\dagger}+a\right)\left(J_{1-}-J_{1+}\right),
\end{equation}

\begin{equation}
\dot{J_{2z}}=\frac{i \lambda_{2}}{\sqrt{N_{2}}}\left(a^{\dagger}+a\right)\left(J_{2-}-J_{2+}\right),
\end{equation}

\begin{equation}
\dot{J_{1-}}=-i \omega_{1}J_{1-}+\frac{2i \lambda_{1}}{\sqrt{N_{1}}}\left(a^{\dagger}+a\right)J_{1z},
\end{equation}

\begin{equation}
\dot{J_{2-}}=-i \omega_{2}J_{2-}+\frac{2i \lambda_{2}}{\sqrt{N_{2}}}\left(a^{\dagger}+a\right)J_{2z},
\end{equation}

\begin{eqnarray}
\dot{a}&=&-\left(\kappa+i \omega_{c}\right)a-\frac{i \lambda_{1}}{\sqrt{N_{1}}}\left(J_{1+}+J_{1-}\right)\\ \nonumber
&-&\frac{i \lambda_{2}}{\sqrt{N_{2}}}\left(J_{2+}+J_{2-}\right).
\end{eqnarray}

Here $\kappa$ is the decay rate of the cavity photons. In addition the magnitude of pseudo-angular momentum is conserved,

\begin{equation}
J_{1z}^{2}+|J_{1-}|^{2}=\frac{N_{1}^{2}}{4},
\end{equation}

\begin{equation}
J_{2z}^{2}+|J_{2-}|^{2}=\frac{N_{2}^{2}}{4}.
\end{equation}

The long time steady state solutions from the equations of motion can be studied with $\dot{\vec{J}}_{i}=0(i=1,2)$ and $\dot{a}=0$. These fixed point solutions can be stable or unstable. Separating $a=a_{1}+i a_{2}$, $J_{i}^{\pm}=J_{ix}\pm J_{iy}$ $(i=1,2)$, one obtains the steady state equations as

\begin{equation}\label{S1}
\kappa a_{1}-\omega_{c}a=0,
\end{equation}

\begin{equation}
\kappa a_{2}+\omega_{c} a_{1}=-\frac{2 \lambda_{1}}{\sqrt{N_{1}}}J_{1x}-\frac{2 \lambda_{2}}{\sqrt{N_{2}}}J_{2x},
\end{equation}

\begin{equation}
\omega_{1} J_{1y}=0,
\end{equation}

\begin{equation}
\omega_{1}J_{1x}=\frac{4 \lambda_{1}}{\sqrt{N_{1}}}a_{1}J_{1z},
\end{equation}

\begin{equation}
\omega_{2} J_{2y}=0,
\end{equation}

\begin{equation}\label{S6}
\omega_{2}J_{2x}=\frac{4 \lambda_{2}}{\sqrt{N_{2}}}a_{1}J_{2z}.
\end{equation}

An analysis of these equations leads us to four types of steady states, namely $(a=0,J_{1z}=\pm N_{1}/2, J_{2z}=\pm N_{2}/2)$. The state $(a=0,J_{1z}=- N_{1}/2, J_{2z}=- N_{2}/2)$ is the normal phase while $(a=0,J_{1z}=N_{1}/2, J_{2z}=N_{2}/2)$ is the inverted phase. The states $(a=0,J_{1z}=-N_{1}/2, J_{2z}=N_{2}/2)$ and
$(a=0,J_{1z}=N_{1}/2, J_{2z}=-N_{2}/2)$ are called mixed phases. As we shall show later that these mixed phases generate interesting non-equilibrium phase diagrams. The critical coupling strength corresponding to the onset of superradiance starting from the normal, inverted or mixed phase is obtained by putting $\vec{J_{i}}=(0,0,\pm N_{i}/2)$ $(i=1,2)$.

This leads us to the following possible critical constants

\begin{equation}
J_{1z}=-\frac{N_{1}}{2}; J_{2z}=-\frac{N_{2}}{2} \left(Normal \\\ Phase\right): \\ \nonumber
\end{equation}

\begin{equation}\label{N1}
\lambda_{1c}=\sqrt{\frac{(\kappa^{2}+\omega^{2})\omega_{1}}{4 \omega}-\frac{\lambda_{2}^{2}\omega_{1}}{\omega_{2}}},
\end{equation}

\begin{equation}
\lambda_{2c}=\sqrt{\frac{(\kappa^{2}+\omega^{2})\omega_{2}}{4 \omega}-\frac{\lambda_{1}^{2}\omega_{2}}{\omega_{1}}},
\end{equation}

\begin{equation}
J_{1z}=\frac{N_{1}}{2}; J_{2z}=\frac{N_{2}}{2} \left(Inverted \\\ Phase\right) : \\ \nonumber
\end{equation}

\begin{equation}
\lambda_{1c}=-\sqrt{\frac{(\kappa^{2}+\omega^{2})\omega_{1}}{4 \omega}+\frac{\lambda_{2}^{2}\omega_{1}}{\omega_{2}}},
\end{equation}

\begin{equation}
\lambda_{2c}=-\sqrt{\frac{(\kappa^{2}+\omega^{2})\omega_{2}}{4 \omega}+\frac{\lambda_{1}^{2}\omega_{2}}{\omega_{1}}},
\end{equation}

\begin{equation}
J_{1z}=- \frac{N_{1}}{2}; J_{2z}=\frac{N_{2}}{2} \left(Mixed \\\ Phase \\\ 1\right) : \\ \nonumber
\end{equation}

\begin{equation}
\lambda_{1c}=\sqrt{\frac{(\kappa^{2}+\omega^{2})\omega_{1}}{4 \omega}+\frac{\lambda_{2}^{2}\omega_{1}}{\omega_{2}}},
\end{equation}

\begin{equation}
\lambda_{2c}=-\sqrt{\frac{(\kappa^{2}+\omega^{2})\omega_{2}}{4 \omega}-\frac{\lambda_{1}^{2}\omega_{2}}{\omega_{1}}},
\end{equation}

\begin{equation}
J_{1z}=\frac{N_{1}}{2}; J_{2z}=-\frac{N_{2}}{2} \left(Mixed \\\ Phase \\\ 2\right) : \\ \nonumber
\end{equation}

\begin{equation}
\lambda_{1c}=-\sqrt{\frac{(\kappa^{2}+\omega^{2})\omega_{1}}{4 \omega}-\frac{\lambda_{2}^{2}\omega_{1}}{\omega_{2}}},
\end{equation}

\begin{equation}\label{M2}
\lambda_{2c}=\sqrt{\frac{(\kappa^{2}+\omega^{2})\omega_{2}}{4 \omega}+\frac{\lambda_{1}^{2}\omega_{2}}{\omega_{1}}}.
\end{equation}

\begin{figure}[h]
\hspace{-0.0cm}
\includegraphics [scale=0.65]{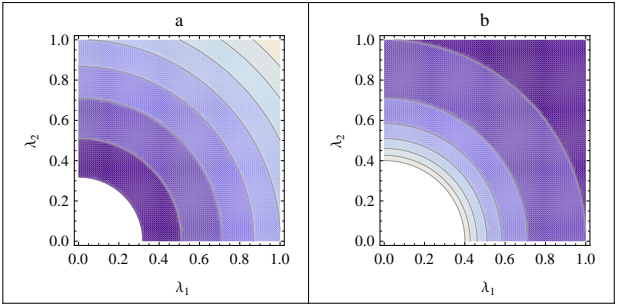}
\caption{(color online)Dynamical phase diagrams of the stable roots corresponding to the normal phase (Eqns.\ref{o1}) in the $(\lambda_{1},\lambda_{2})$ plane. The parameters used are $\omega_{1}/\kappa$ $=$  $\omega_{2}/\kappa$ $=$ $\omega_{c}/\kappa$. The white region is the non-superradiant normal phase while the superradiant phase is indicated by the contours. The darker region in the contour corresponds to low superradiance while light region corresponds to high superradiance. The left plot is the $\omega_{+}$ root while the right plot is the $\omega_{-}$ root.}
\end{figure}\label{fig2}

These set of expressions reveals one interesting point that the critical coupling strength for one set of atoms depends on the coupling strength of the other set of atoms. For a given set of $J_{1z}$ and $J_{2z}$,  Eqns. \ref{N1}-\ref{M2} determines the boundary between the nonsuperradiant (normal/inverted/mixed) and superradiant phase.  A trivial manipulation of Eqns. \ref{S1}-\ref{S6} leads us to the following equation for $J_{1z}=-N_{1}/2$ and $J_{1x}=0$,

\begin{equation}
J_{2x}\left(\omega_{2}\left(\kappa^{2}+\omega_{c}^{2}\right)+\frac{8 \lambda_{2}^{2}}{N_{2}}\omega_{c} J_{2z} \right)=0.
\end{equation}

Now there are two possibilities depending on whether $J_{2x}=0$ and $J_{2z}=\pm N_{2}/2$ or $J_{2x}\neq 0$ and $J_{2z}=-N_{2}\omega_{2} (\kappa^{2}+\omega_{c}^{2})/8 \lambda_{2}^{2} \omega_{c}$. The first condition implies that both the set of atoms are in the non-superradiant phase. The second solution corresponds to the case where the first set of atoms are in the non-superradiant normal phase while the second set of atoms are in the superradiant phase. Substituting the second expression for $J_{2z}$ from above in the expression for $\lambda_{1c}=\sqrt{\frac{\omega_{1}(\kappa^{2}+\omega_{c}^{2})}{4 \omega_{c}}+\frac{2 \lambda_{2}^{2} \omega_{1}}{N_{2} \omega_{2}}J_{2z}}$,one obtains $\lambda_{1c}=0$. This implies that by keeping one coupling strength arbitrarily low, one could enter the superradiant phase by manipulating the second coupling strength alone. This point would be more clear when we discuss the dynamical phase diagrams in the next section.

\section{Dynamical Phase Diagrams}

In this section, we explore the fluctuation dynamics above the steady state (fixed points). In particular, we will consider the instability of the normal ($\downarrow \downarrow$), inverted ($\uparrow \uparrow$)and mixed phases ($\uparrow \downarrow$ or $\downarrow \uparrow$). To this end, we write $a=a_{0}+\delta a$, $J_{i-}=J_{i-}^{0}+\delta J_{i-}$ $(i=1,2)$, where $a_{0}=0$, $J_{i-}=0$ and $J_{iz}=\pm N_{i}/2$ $(i=1,2)$. Substituting these into Eqns.(4)-(6), one obtains the linearized equations

\begin{eqnarray}
\dot{\delta a}&=& -\left(\kappa+i \omega_{c}\right) \delta a-\frac{i \lambda_{1}}{\sqrt{N_{1}}} \left( \delta J_{1+}+ \delta J_{1-}\right)\\ \nonumber
&-& i\frac{i \lambda_{2}}{\sqrt{N_{2}}} \left( \delta J_{2+}+ \delta J_{2-}\right)
\end{eqnarray}

\begin{equation}
\dot{\delta J_{1-}}=-i \omega_{1} \delta J_{1-}+ \frac{2 i \lambda_{1}}{\sqrt{N_{1}}} \left( \delta a^{\dagger}+\delta a\right) J_{1z},
\end{equation}

\begin{equation}
\dot{\delta J_{2-}}=-i \omega_{2} \delta J_{2-}+ \frac{2 i \lambda_{2}}{\sqrt{N_{2}}} \left( \delta a^{\dagger}+\delta a\right) J_{2z}.
\end{equation}

\begin{figure}[h]
\hspace{-0.0cm}
\includegraphics [scale=0.65]{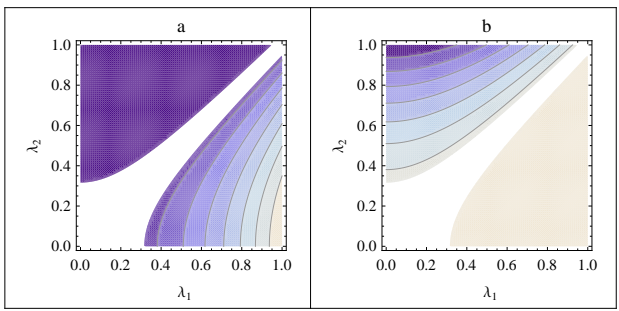}
\caption{(color online)Dynamical phase diagrams of the stable roots corresponding to the mixed phase $1$ (Eqns.\ref{o3}) in the $(\lambda_{1},\lambda_{2})$ plane. The parameters used are $\omega_{1}/\kappa$ $=$  $\omega_{2}/\kappa$ $=$ $\omega_{c}/\kappa$. The left plot is the $\omega_{+}$ root while the right plot is the $\omega_{-}$ root. The white region is the non-superradiant mixed phase $1$ while the superradiant phase is indicated by the contours.}
\end{figure}\label{fig3}

We now write $\delta a= A e^{-i \eta t}+B^{*} e^{i \eta^{*} t}$, $\delta J_{1-}=C e^{-i \eta t}+ D^{*} e^{i \eta^{*} t}$ and $\delta J_{2-}=E e^{- i \eta t}+ F^{*} e^{i \eta^{*} t}$ and equate coefficients with the same time dependence to obtain algebraic equations for $A$, $B$, $C$, $D$, $E$ and $F$. The corresponding self consistency equations yields a quadratic equation for $\omega$, whose roots characterize the possible instabilities for $\eta =0$. These instabilities describe the boundaries in the emerging dynamical phase diagrams. In particular, the various boundaries between exponentially growing and decaying fluctuations are given as:

\bigskip
\bigskip

Normal Phase:

\begin{equation}\label{o1}
\omega_{\pm}=2\left(\frac{\lambda_{1}^{2}}{\omega_{1}}+\frac{\lambda_{2}^{2}}{\omega_{2}}\right)\pm \sqrt{4\left(\frac{\lambda_{1}^{2}}{\omega_{1}}+\frac{\lambda_{2}^{2}}{\omega_{2}}\right)^2-\kappa^{2}},
\end{equation}

Inverted Phase:

\begin{equation}\label{o2}
\omega_{\pm}=-2\left(\frac{\lambda_{1}^{2}}{\omega_{1}}+\frac{\lambda_{2}^{2}}{\omega_{2}}\right)\pm \sqrt{4\left(\frac{\lambda_{1}^{2}}{\omega_{1}}+\frac{\lambda_{2}^{2}}{\omega_{2}}\right)^2-\kappa^{2}},
\end{equation}

Mixed Phase 1:

\begin{equation}\label{o3}
\omega_{\pm}=2\left(\frac{\lambda_{1}^{2}}{\omega_{1}}-\frac{\lambda_{2}^{2}}{\omega_{2}}\right)\pm \sqrt{4\left(\frac{\lambda_{1}^{2}}{\omega_{1}}-\frac{\lambda_{2}^{2}}{\omega_{2}}\right)^2-\kappa^{2}},
\end{equation}

Mixed Phase 2:

\begin{equation}\label{o4}
\omega_{\pm}=-2\left(\frac{\lambda_{1}^{2}}{\omega_{1}}-\frac{\lambda_{2}^{2}}{\omega_{2}}\right)\pm \sqrt{4\left(\frac{\lambda_{1}^{2}}{\omega_{1}}-\frac{\lambda_{2}^{2}}{\omega_{2}}\right)^2-\kappa^{2}},
\end{equation}

Note that the "inverted phase" is the inversion of the "normal phase" around $\omega=0$ boundary while "mixed phase $2$" is the mirror inversion of "mixed phase $1$". The contour plot of the stable roots $\omega_{\pm}$ of Eqns.\ref{o1} as a function of $\lambda_{1}$ and $\lambda_{2}$ for the normal phase is shown in Fig.2. Fig.2(a) shows the boundary separating the non-superradiant phase and the superradiant phase for the $\omega_{+}$ root. This boundary is the curve that joins $\lambda_{1c}$ (with $\lambda_{2}=0$) and $\lambda_{2c}$ (with $\lambda_{1}=0$). Below this boundary is the non-superradiant normal phase while above this curve is the superradiant phase. As we move along the $y$-axis $(\lambda_{1}=0)$, we reach the superradiant phase at $\lambda_{2c}=(\kappa^{2}+\omega^{2})\omega_{2}/4\omega$. This analysis agrees with our steady state analysis of the previous section. Thus along the $x$ or the $y$ axis, the system behaves as if only one set of atoms are present. In any other direction, both set of atoms contribute to the dynamics. Note that white region is the non-superradiant normal phase while the superradiant phase is indicated by the contours. The darker region in the contour corresponds to low superradiance while light region corresponds to high superradiance. Fig2b shows the plot of $\omega_{-}$ root. The combination of $\omega_{+}$ and $\omega_{-}$ determine the complete boundary between the non-superradiant phase and superradiant phase. The phase diagrams of the inverted phase (not shown) is the mirror inversion of the normal phase.
In a similar manner, one can determine the dynamical phase diagrams for the mixed phases. In fig.3(a) and 3(b), we demonstrate this for the mixed phase $1$. A new kind of dynamical phase diagram emerges for the mixed phase. The phase diagram now splits into two distinct superradiant regimes separated by the non-superradiant phase. The two critical points $\lambda_{1c}$ $(\lambda_{2}=0)$ and $\lambda_{2c}$ $(\lambda_{1}=0)$ along the $x$ and $y$ axis are still the same. In Fig.3a, $\omega_{+}$ root is shown and on moving along the $x$ axis (increasing $\lambda_{1}$), we encounter the usual superradiant phase with increasing energy. On the other hand, moving along the $y$ axis (increasing $\lambda_{2}$), we get a superradiant phase of constant low energy. There are regions in the phase diagram, where even when $\lambda_{1}>\lambda_{1c}$ and $\lambda_{2}>\lambda_{2c}$, the system stays in the non--superradiant phase. Interestingly for $\lambda_{1}=\lambda_{2}$, the superradiant phase can never be reached. Infact the energy landscape in the $(\lambda_{1},\lambda_{2})$ plane gives an impression of anti-crossing of energy levels. Fig.3(b) shows the plot of $\omega_{-}$ root whose behavior is opposite to that of the $\omega_{+}$ root. A superradiant phase with decreasing energy is encountered along the $y$ axis while a constant high energy phase is encountered along the $x$ axis. Note that if we choose $\omega_{1}\neq \omega_{2}$, then the energy plots of Fig.2 and Fig.3 becomes asymmetric (figure not shown).

The current predictions can be tested in an experiment similar to that of Ref.\citep{bau1} but with two species condensate. In addition one has to look into the long duration of these experiments beyond the $10$ $ms$  time scale as noted in Ref.\citep{bhaseen}

\bigskip

\section{conclusions}

In summary, we have investigated the non-equilibrium quantum phases of a two-atom Dicke model, which is realized in a collection two set of cold atoms coupling simultaneously to a single quantized cavity mode. Within the framework of the non-equilibrium two-atom Dicke model, we reveal a rich and new set of phase diagrams. We have shown the existence of a new kind of quantum phase transition from the non-superradiant mixed phase (where one set of atoms are in the normal phase while the other set is in the inverted phase) to the superradiant phase. In addition, we have demonstrated that in the quantum phase diagram of the mixed phase, there are regions where the superradiant phase cannot exist even if the light-matter coupling constants of both set of atoms are above the critical value. These predictions can be realized in a two species cold atoms in an optical cavity.

\section{Acknowledgements}
 A. Bhattacherjee acknowledges financial support from the Department of Science and Technology, New Delhi for financial assistance vide grant SR/S2/LOP-0034/2010.

\end{document}